\documentclass{aa}
\usepackage{graphicx}
\usepackage{natbib}
\def\mspace{\hspace*{1 cm}} 

\def\arcsec{\hbox{$^{\prime\prime}$}}
\def\utw{\smash{\rlap{\lower5pt\hbox{$\sim$}}}}
\def\udtw{\smash{\rlap{\lower6pt\hbox{$\approx$}}}}

\def\diameter{{\ifmmode\mathchoice
{\ooalign{\hfil\hbox{$\displaystyle/$}\hfil\crcr
{\hbox{$\displaystyle\mathchar"20D$}}}}
{\ooalign{\hfil\hbox{$\textstyle/$}\hfil\crcr
{\hbox{$\textstyle\mathchar"20D$}}}}
{\ooalign{\hfil\hbox{$\scriptstyle/$}\hfil\crcr
{\hbox{$\scriptstyle\mathchar"20D$}}}}
{\ooalign{\hfil\hbox{$\scriptscriptstyle/$}\hfil\crcr
{\hbox{$\scriptscriptstyle\mathchar"20D$}}}}
\else{\ooalign{\hfil/\hfil\crcr\mathhexbox20D}}%
\fi}}

\bibpunct{(}{)}{;}{a}{}{,}
\begin{document}
\title{The spectral-curvature parameter: an alternative
 tool for the analysis of synchrotron spectra}
\author{B.W. Sohn\inst{1,}\inst{2} \and U. Klein\inst{2}
\and K.-H. Mack\inst{3,}\inst{4,}\inst{2}}
\offprints{B.W. Sohn}
\institute{Max-Planck-Institut f\"ur Radioastronomie, Auf dem H\"ugel 69, 
D-53121 Bonn, Germany
\and
Radioastronomisches Institut der Universit\"at Bonn,
Auf dem H\"ugel 71, D-53121 Bonn, Germany
\and
ASTRON/NFRA, Postbus 2, NL-7990 AA Dwingeloo, The Netherlands
\and
Istituto di Radioastronomia del CNR, Via P. Gobetti 101, I-40129 Bologna,
Italy}
\date{}
%
\abstract{
A new intuitive tool for the analysis of synchrotron spectra is presented.
The so--called {\it Spectral Curvature Parameter} (SCP), when plotted versus
the high--frequency spectral index ($\alpha_{\rm high}$) of synchrotron
sources, provides crucial parameters on the continuum spectrum of
synchrotron radiation without the more complex modeling of spectral
ageing scenarios.
An important merit of the SCP-$\alpha$ diagram, in respect to the conventional
colour-colour diagram (i.e. $\alpha$-$\alpha$ diagram), is the enhanced 
reliability of extracting multiple injection spectra, $\alpha_{\rm inj}$.
Different from the colour-colour diagram, tracks of different 
$\alpha_{\rm inj}$s, 
especially when the synchrotron particles are young, exhibit less overlap and
less smearing in the SCP-$\alpha$ diagram. 
Three giant radio galaxies (GRGs) and a sample of Compact steep spectrum
(CSS) souces, which are parti\-cularly suitable for
this kind of analysis, are presented.
GRGs exhibit asymmetries of their injection spectral indices
$\alpha_{\rm inj}$ in the SCP - $\alpha_{\rm high}$ diagram.
The obtained $\alpha_{\rm inj}$s and the trends in the sources are 
cross-checked with the literature and show remarkable confidence.
Besides the spectral steepening which is well understood in the
framework of synchrotron ageing models, spectral flattening is prominent in 
the radio lobes. The spectral flattening is a clue to efficient 
re-acceleration processes in the lobes. This implies that interaction with 
the surrounding intergalactic or intra-cluster medium is an important 
characteristic of GRGs. In the SW lobe of DA\,240, there is a clear sign of 
CI and KP/JP bifurcation at the source extremity. This indicates a highly 
relativistic energy transportation from the core or {\it in situ} acceleration
in this typical FR I lobe. Our analysis proves, if exists, KP spectra imply 
the existence of strong $B_{\rm sync}$ field with $B_{\rm sync} > B_{\rm CMB}$.
In the CSS sources, our result confirms the CI model and 
$B_{\rm sync} \gg B_{\rm CMB}$. The synchrotron self-absorption is significant 
in the CSS sample.
}
\authorrunning{Sohn et al.}
\titlerunning{SCP - $\alpha$ diagram, an alternative tool}
\maketitle
%
%
\section{Introduction}

Spectra of synchrotron sources from the radio to the X-ray regime reflect the
energy distribution of relativistic particles, i.e. electrons whose energy
distributions obey a power-law. In general, the synchrotron emissivity also
follows a power-law \citep{pacholczyk:1970}. While the conventional spectral 
indices only provide the spectral slope between the two observing frequencies,
a multi-frequency data set can also disclose spectral curvatures over a larger
frequency range. The significance of the shape of synchrotron spectra has
been underlined early on by \citet{kardashev:1962, pacholczyk:1970, 
pacholczyk:1977, jaffe:1973}, who were among the first to describe and apply
synchrotron loss models to flux densities obtained at se\-veral frequencies.

It is obvious that the information on the spectral shape of a source under
the effects of ageing, adiabatic expansion etc. provides an important tool
for understanding source evolution. If the injection of relativistic 
particles following a power-law is restricted to a certain region -- the 
cores and/or hot spots of radio galaxies -- and if the observation is 
performed with appropriate resolution, one can detect regional variations of 
spectral curvature by means of the above-mentioned physical processes. 
Since synchrotron and Inverse Compton losses are the main energy dissipation 
processes in radio galaxies, in particular at high and intermediate radio 
frequencies ($> 1$ GHz), large efforts have been made to explain the 
variation of spectral curvatures -- often by modeling two-frequency data -- 
of radio galaxies with the synchrotron ageing theory 
\citep[e.g.][]{alexander:1987, alexander2:1987,klein:1995, feretti:1998, 
murgia:1999}. A proper determination of parameters like the injection 
spectral index $\alpha_{\rm inj}$ (the spectrum of the electron distribution 
immediately after acceleration, $N(E) \propto E^{-p}$, 
$\alpha_{\rm inj} = (p - 1)/2 $ or the break frequency $\nu_{\rm br}$ 
, the frequency at which spectral steepening occurs can be obtained with a 
spectral ageing analysis \citep[e.g.][]{carilli:1991, mack:1998, murgia:1999}.
This requires, however, the fitting of appropriate models with several 
parameters, thus high-quality measurements at many fequencies with a good 
signal-to-noise ratio are essential. 

In order to fit synchrotron and Inverse-Compton losses, three models are 
widely used: The continuous injection (CI) model \citep{pacholczyk:1970} 
assumes a mixture of electron populations of various synchrotron ages. 
In this model, permanent replenishment of fresh particles is assumed so that 
the injection spectral index steepens to its final value of 
$\alpha_{\rm inj} + 0.5 $ beyond the break frequency. 
The Kardashev-Pacholczyk (KP) model \citep{kardashev:1962, pacholczyk:1970} 
merely includes a single injection of power-law distributed electrons. 
The pitch angles of the electrons are assumed to be constant with time. 
The high-frequency slope in this model is $\frac{4}{3} \, \alpha_{\rm inj}+1$. 
The Jaffe-Perola (JP) model \citep{jaffe:1973} incorporates 
-- similar to KP -- a single injection but permits permanent pitch angle 
isotropization. Beyond the break frequency this model leads to an exponential 
steepening of the high-frequency spectrum. A sketch of the different tracks 
of the various ageing models in the classical $\log(S)-\log(\nu)$ space can 
be found in the work of \citet[][see their Fig.~1]{carilli:1991}.

In many cases the spectral ageing analysis yielded significant results. 
However, high-resolution multi-frequency studies of two prototypical nearby 
radio galaxies  -- 3C\,449 (FR I type) by \citet{katz-ston:1997} and 
Cygnus~A (FR~ II type) by \citet{carilli:1991}, show trends that cannot be 
explained by the synchrotron ageing theory alone. The first problem is that 
jets and lobes (3C\,449), and hot-spots and lobes (Cygnus~A) have different 
injection spectra. The second problem to be dealt with is that of the 
microscopic physical conditions. While possible physical conditions such as 
turbulent magnetic fields and inverse-Compton scattering by cosmic microwave 
background photons favour the pitch-angle isotropized (JP) model, the 
observational results appear to support the constant pitch angle (KP) model. 
This could mean that the nature of the spectral curvature 
is more complex than expected from the synchrotron ageing theory alone. 

\citet{carilli:1996} have pointed out the necessity of an appropriate
empirical analysis that is not tied to any theoretical model in order to 
find the real trends in sources. Here we present a new method which can 
fulfill this requirement. It also aims at a quick determination of the 
injection spectral index and the best suited model to fit the observed 
spectrum. It fills the gap between the simple spectral index study and 
the much more complex spectral ageing analysis. It is also suited to 
provide first guesses of the parameters to be fit in a spectral ageing 
analysis, therefore making the fit procedure less susceptible to local 
minima in the error space.
%
%
\section{The spectral curvature parameter-$\alpha$ diagram}
This method is based on the so-called spectral curvature parameter (SCP).
It is defined as\\
\mspace $SCP \equiv \frac{\alpha_{\rm high} - \alpha_{\rm low}}
{\alpha_{\rm high} + \alpha_{\rm low}}$,\vspace{3mm}\\ 
When displayed as a function of the spectral index $\alpha$ with\\
\mspace $I_{\nu} \propto \nu^{- \alpha}$,\vspace{3mm}\\
the SCP indicates how the spectrum evolves, starting from its pure power-law. 
As $\alpha_{\rm high}$ is more sensitive to both spectral steepening and 
spectral flattening than $\alpha_{\rm low}$ 
\citep{pacholczyk:1970, eilek:1991, carilli:1991},
we employ $\alpha_{\rm high}$ as the counter axis of SCP. Though the classical
$log(S_{\nu})$ - $log(\nu)$ diagram is the best way to test ageing models
for a single-spectrum population, it is not a straight-forward tool to unveil
different trends in an extended source. It is here where the SCP-$\alpha$
diagram has its power. Each spectrum is represented by a point in the
SCP-$\alpha$ plane, and a lot of spectra from an extended source can be
drawn in this plane.
\begin{figure}[tbp]
  \resizebox{\hsize}{!}{\includegraphics{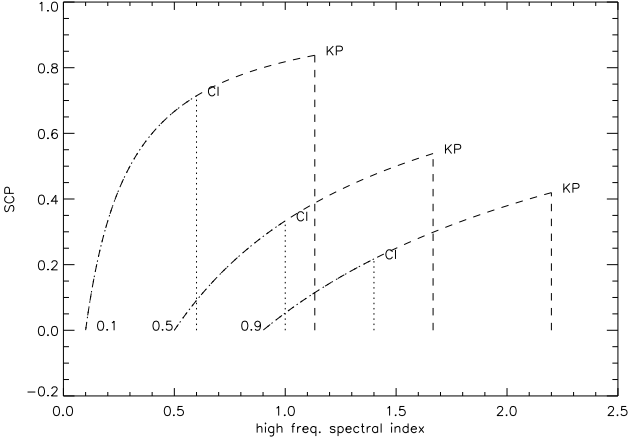}}
  \caption[Schematic SCP--$\alpha_{high}$ diagram]{Schematic SCP--
$\alpha_{high}$ diagram of power-law injection spectra, 
$\alpha_{\rm inj} = 0.1, 0.5, 0.9$, undergoing synchrotron losses. The numbers 
represent the injection spectral index of each line. Along the dash-dotted 
curves, all models, i.e. CI, KP and JP are possible. At the maximum SCP values 
of the CI models the tracks between CI (dotted lines) and KP/JP (dashed lines) 
split at the points marked 'CI'. Along the dashed curves the KP and JP models 
are possible. At SCP$_{\rm max}$ of the KP model(marked 'KP') the KP (dashed 
straight lines) and JP models (imaginary curves approaching SCP $= 1$) take 
separate tracks. The break frequency, $\nu_{\rm br}$, reaches the low frequency
regime such that $\nu_{1} \ge \nu_{\rm br} \ge \nu_{2}$ where 
$\nu_{1}, \nu_{2}$ are the frequencies used for the estimation of 
$\alpha_{\rm low}$. The KP spectra will fall vertically to $SCP = 0$, since 
$ \alpha_{\rm high} = \alpha_{\rm KP, br} = 4/3 \alpha_{\rm inj} + 1$. The JP 
high-frequency tail falls off exponentially beyond the break frequency in the 
$log(S_{\nu})$ - $log(\nu)$ diagram, therefore the track will approach 
asymptotically SCP = 1 in the SCP-$\alpha_{\rm high}$ regime. Tracks of these 
most common synchrotron ageing models in the $log(S_{\nu})$ - $log(\nu)$ 
parameter space are sketched by \citet{carilli:1991}.}
  \label{fig:scp}
\end{figure}

Fig.~\ref{fig:scp} illustrates the schematic tracks of the power-law spectra 
undergoing synchrotron ageing. More realistic simulations of SCP--$\alpha$ 
diagram including the Inverse Compton equivalent field of Cosmic Microwave 
Background radiation, i.e. $B_{\rm CMB}$, are presented in what follows. 
The dash-dotted curved lines represent tracks where the break frequency 
has not yet reached the low-frequency regime (i.e. where $\alpha_{\rm low}$ is 
determined). The CI, KP and JP models produce different SCP ranges. This makes 
it easy to distinguish between the different models in the 
SCP-$\alpha_{\rm high}$ plane. Different injection spectral indices also 
follow different tracks.

Since both the CI and the KP model predict a power-law spectrum also beyond
the break frequency, namely the so-called broken power-law \citep{eilek:1991},
we can calculate the maximum SCPs in these cases. For $\alpha_{\rm inj} = 0.5$,
these are $SCP_{\rm CI, max} \sim 0.33$ and $SCP_{\rm KP, max} \sim 0.54$. 
In contrast, the high-frequency part of the JP model has a non-power-law 
curvature, viz. an exponential one. Therefore, the tracks of JP spectra 
asymptotically approach SCP$ = 1.0$. In any case, 
$\alpha_{\rm h} > (\frac{4}{3}) \alpha_{\rm inj} + 1$ is predicted by the JP 
model only.

%
\begin{figure}[tbp]
  \resizebox{\hsize}{!}{\includegraphics{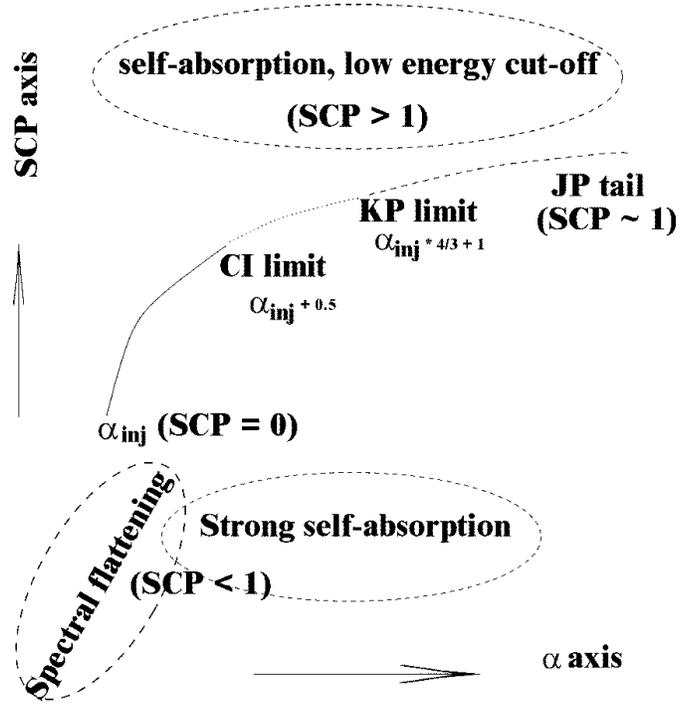}}
  \caption[Schematic diagram of a synchrotron source]{Schematic diagram of a 
synchrotron source with a given injection spectrum and with various processes 
that affect the spectral curvature. In this sketch, the classical pitch angle 
models, i.e. KP and JP are considered.}
  \label{fig:sketch}
\end{figure}

\subsection{SCP-$\alpha$ diagram and colour-colour diagram}
\begin{figure}[tbp]
\resizebox{\hsize}{!}{\includegraphics{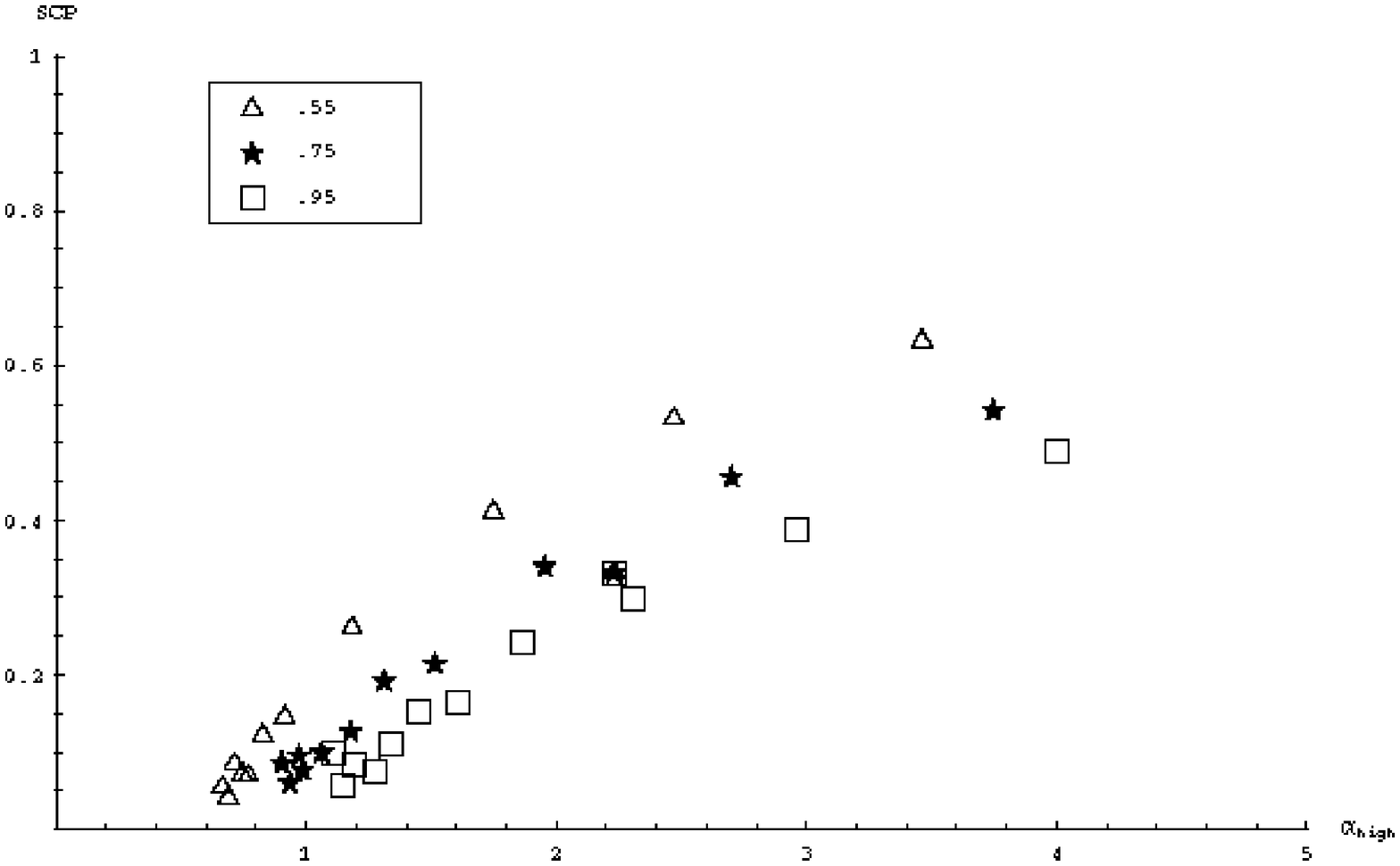}}
\resizebox{\hsize}{!}{\includegraphics{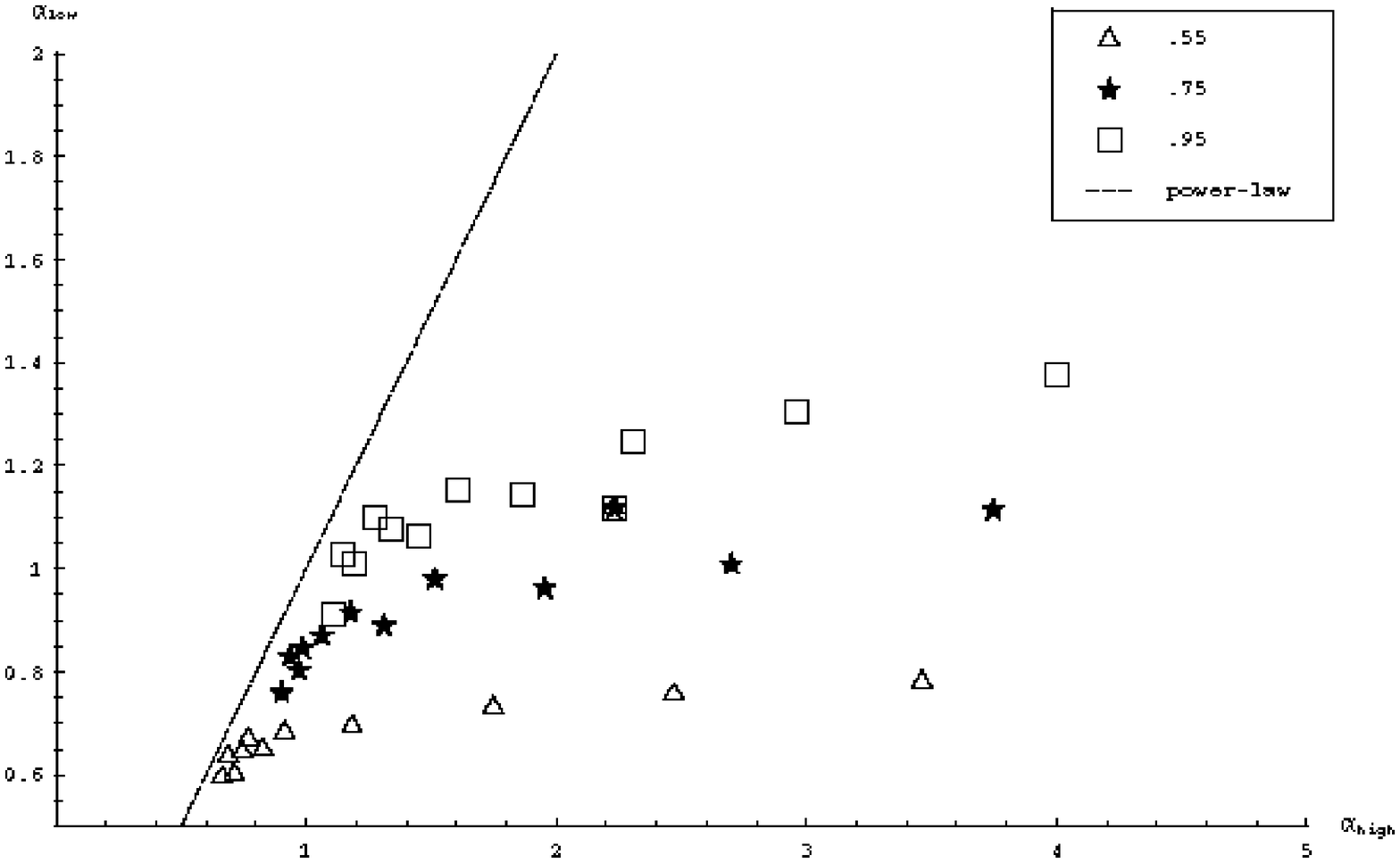}}
  \caption[Comparison between SCP-$\alpha$ and the Colour-Colour diagram, JP 
model]{Comparison between SCP-$\alpha$ and the colour-colour diagram with 
$B_{\rm sync} = 6.0 \mu G$, $B_{\rm CMB} = 3.2 \mu G$ and JP model. 
The two field strengths are simply additive in JP model, since they both are
isotropic. The frequency intervals were selected to be similar to standard 
observing frequencies in the classical radio regime. Open triangles are the 
result with $\alpha_{\rm inj} = 0.55$. Filled stars 
are the result with $\alpha_{\rm inj} = 0.75$. Open squares are the result with
 $\alpha_{\rm inj} = 0.95$. The dotted line in the Colour-Colour diagram is 
$\alpha_{\rm low} = \alpha_{\rm high}$ line, i.e. a pure power-law while in the
SCP-$\alpha$ diagram this corresponds to SCP = 0.
In the simulation, we assumed the low energy cut-off, $E_0$, and the high 
energy cut-off by CMB photons. The energy range is binned in 200 bins.
At given frequencies, the synchrotron radiation contribution from each bin is
calculated.
}
  \label{fig:6G_jp}
\end{figure}

An advantage of the SCP-$\alpha$ diagram with respect to the colour-colour
diagram \citep{katz-ston:1993}, is seen when the source has multiple 
$\alpha_{\rm inj}$s. As predicted in the particle ageing theory, the track of 
an aged $\alpha_{\rm inj}$ do not leave immediately the 
$\alpha_{\rm low} = \alpha_{\rm high}$ line (see Fig.~\ref{fig:6G_jp}). 
Because of the overlap of the $\alpha_{\rm low} = 
\alpha_{\rm high}$ line and the spread of points parallel to it in the 
colour-colourdiagram (hereafter C-C diagram), trends in the source can not be 
easily identified. We emphasize that there are model dependent aspects 
in the spectral tomography or the classical synchrotron ageing analysis, 
\citep[e.g.][]{alexander:1987, murgia:1999}.

For example, the synchrotron ageing analysis can hardly achieve a 
pixel-to-pixel study. The synchrotron ageing analysis is therefore done in a 
way with certain subdivided integrated areas. As a property of the 
integration, the obtained $\alpha_{\rm inj}$ is strongly biased by bright 
structures in these areas. The spectral tomography aimed at solving this 
problem. This technique isolates from an assumed or known $\alpha_{\rm inj}$ 
component a 'different' component. The $\alpha_{\rm inj}$ of this 'different' 
component is not directly obtained in the spectral tomography. 
The spectral tomography with multiple $\alpha_{\rm inj}$s could be much more 
complex than the classical synchrotron ageing analysis\citep{katz-ston:1994}.
The tool suggested here will extract 
$\alpha_{\rm inj}$s without the bias due to bright structures and without 
complex tomographical mapping. It can serve as a 'precensor', such as to 
select the area of integration in the synchrotron ageing analysis correctly, 
thus providing quite reliable physical parameters.

The clear bifurcations on the SCP-$\alpha$ diagram, between CI and 
KP/JP and between KP and JP, are further merits of the SCP-$\alpha$ diagram.
On the other hand, the tracks of the different models beyond the $\nu_{\rm br}$
show overlaps on the C-C diagrms.

In both cases, i.e. CI and KP, straight vertical lines arise, while in the 
KP/JP and JP case the original curves are maintained. This fact makes the 
selection of the proper ageing model easier than in the C-C diagram. Of course,
under the influence of CMB, this last argument is only true if $B_{\rm sync}$ 
is (much) stronger than $B_{\rm CMB}$. The weak point of both, the C-C diagram 
and the SCP-$\alpha$ diagram is the loss of positional information of the 
spectra. In order to compensate for this weakness, we present the SCP-$\alpha$ 
diagram and the SCP map together. In this way, the position information of 
spectra can be restored. Some first results of this exercise will be shown in 
the next section.
%
%
%
%
%
\section{Application to radio galaxies}

In this section, we present SCP-$\alpha$ diagrams of three Giant Radio 
Galaxies (GRGs) and of a sample of Compact Steep Spectrum (CSS) sources. The 
integrated spectra have been analyzed by \citet{mack:1998} for GRGs and by 
\citet{murgia:1999} for CSS sources with synchrotron ageing models. Error 
bars in the diagrams are $\sigma_{\rm scp}$ and $\sigma_{\alpha}$. These are 
estimated as shown below,\\
\mspace$\sigma_{\rm scp}^2 = \frac{4(\alpha_{\rm high}^2 + \alpha_{\rm low}^2)}
{( \alpha_{\rm high} + \alpha_{\rm low})^4}  (\alpha_{\rm high}^2 
\sigma_{\alpha_{\rm high}}^2 + \alpha_{\rm low}^2 
\sigma_{\alpha_{\rm low}}^2 )$ \newline
\mspace$\sigma_{\alpha} = 1/ \log (\lambda_2 /\lambda_1) \sqrt{(\sigma_{\rm I, 1}^2 / I_1^2) +(\sigma_{\rm I, 2}^2 / I_2^2)}$.\vspace{3mm}\\
$\alpha_{\rm low}$ and $\alpha_{\rm high}$ are the spectral indices ($I_{\nu}
\propto \nu^{- \alpha}$), obtained at low (e.g. $<$ 1 GHz) and at high (e.g.
$>$ 1 GHz) frequencies, respectively. Since two independent spectral indices 
are needed for the SCP, observations over at least three different 
frequencies are necessary.

\subsection{Giant Radio Galaxies}
These objects are classified by their projected linear sizes. The measurements
used here have been performed by \citet{mack:1997} at frequencies between 
326~MHz and 10.6~GHz. We use four frequencies in our analysis, viz. 326~MHz, 
610~MHz, 4.8~GHz and 10.6~GHz. We compute 
$\alpha^{610 {\rm MHz}}_{326 {\rm MHz}}$ as $\alpha_{\rm low}$, and 
$\alpha^{10.6 {\rm GHz}}_{4.8 {\rm GHz}}$ as $\alpha_{\rm high}$. 
All maps were convolved to a common resolution of $150\arcsec \times 150 
\arcsec$, SCP-$\alpha$ diagrams were produced for brightness levels above 
$\sim 3 \sigma$. In general, the low--frequency spectral indices in the lobes 
of all three sources remain relatively constant, $\alpha_{\rm low} \sim 
\alpha_{\rm inj}$. This means that neither ageing processes nor synchrotron 
self--absorption play an important r\^{o}le at low frequencies in the regions 
of interest. In what follows, we shall discuss the results for the three GRGs 
investigated here in detail. For the best performance, if needed, the cubic 
convolution interpolation method with a value of -0.5 is used when regridding 
\citep{park:1983}. The linear convolution interpolation shows marginal 
difference.

%
%
%
%
%
\subsubsection{DA\,240}
The radio morphology of DA\,240 is symmetric at low frequencies, but becomes
increasingly asymmetric towards higher frequencies \citep{mack:1997}. 
At 326~MHz, DA\,240 is seen as a `Fat Double'. The SW fat lobe has disappeared
at 10.6~GHz, forming an elongated edge-darkened FR-I-type lobe. On the 
contrary, the NE lobe maintains its `fat round' shape up to 10.6~GHz.

\begin{figure*}[tbp]
  \centering\includegraphics[width=12cm]{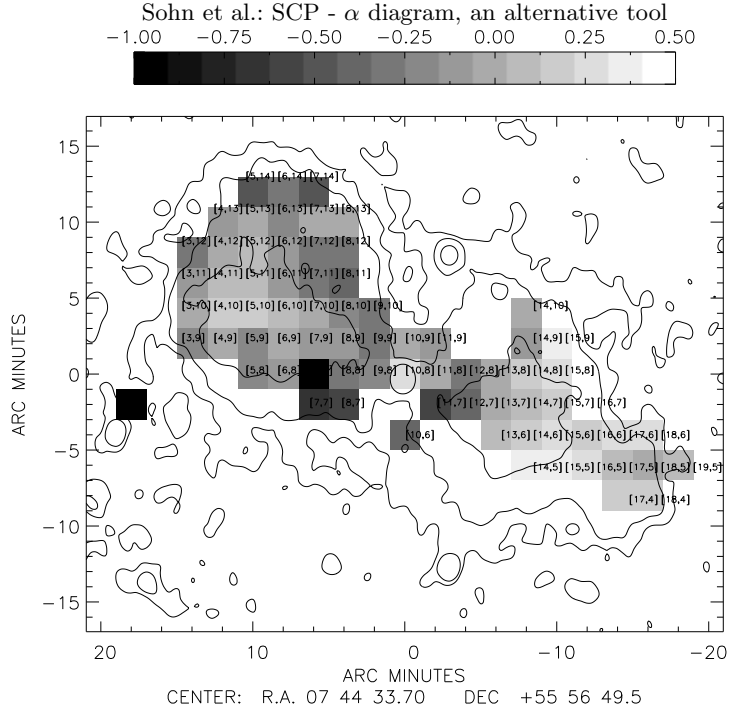}
  \caption[SCP map of DA\,240]{SCP map of DA\,240. 
The contours show the total intensities at 326~MHz \citet{mack:1997}.
Contour levels are 3$\sigma_{\rm I}$, 10$\sigma_{\rm I}$ and 
50$\sigma_{\rm I}$.
The SCP map shows asymmetry. The spectral curvature of the NE lobe is nearly
a power-law in the vicinity of its bright centre [5,10]. The envelope of
the NE lobe shows spectral flattening. This implies an effective {\it in situ}
acceleration, possibly through the interaction with a surrounding thermal 
magneto-ionic medium whose existence is detected via its rotation measure 
(\citet{sohn:2003}). The SCP values of the SW lobe are in a range which is 
expected by the particle ageing theory. The SW lobe has no well-defined 
envelope at high frequencies. Its SW extremity shows spectral flattening with 
respect to its bright centre. The fact that all of DA240's extremities have a 
SCP flatter than their brightness centres implies that the interaction (i.e. 
re-acceleration) with its environment is an essential mechanism for the 
growth of this GRG.
}
  \label{fig:DA240_scp_map}
\end{figure*}

A fit to the SCP-$\alpha$ diagram yields steep injection spectra, 
$\alpha_{\rm inj} \sim 0.82$ (NE lobe) and $\sim 0.94$ (SW lobe). These 
unusually steep and asymmetric injection spectra have already been reported by 
\citet{mack:1998}, viz. 0.76 and 0.97 for the NE and SW lobe respectively. 
Those authors used the synchrotron ageing technique. The diffe\-rence of the 
injection spectral indices is relatively large in the NE lobe, since the 
region with SCP$<0$ of the NE lobe (Fig. \ref{fig:DA240_scp_map}) is included 
in the integrated synchrotron ageing calculation. Including this 
flatter-spectrum region, $\alpha_{\rm high} = 0.5 \ldots$ 0.85 makes the 
synchrotron ageing estimate uncertain. On the whole, our intuitive 
rapid estimation shows good agreement with their result.

\begin{figure*}[tbp]
  \centering\includegraphics[width=12cm]{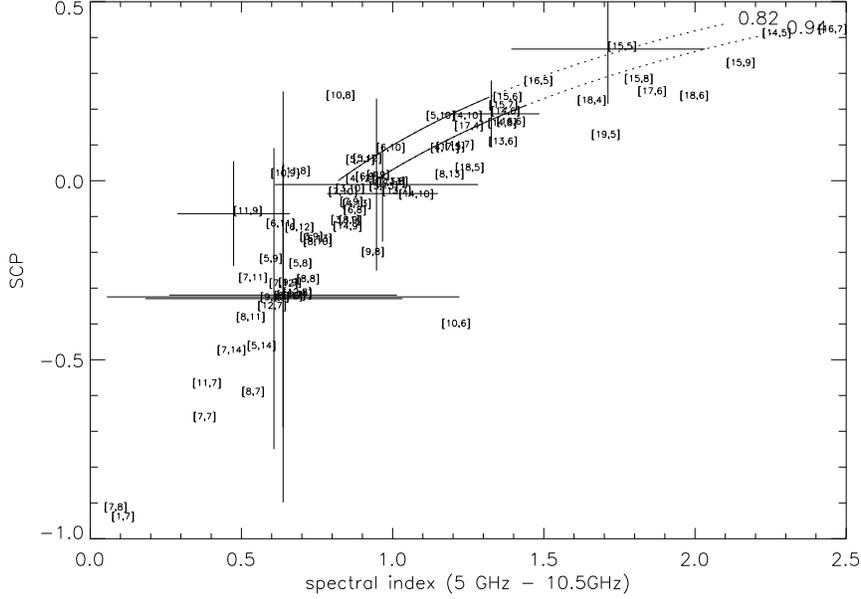}
  \caption[SCP-$\alpha$ diagram of DA\,240]{SCP-$\alpha$ diagram of DA\,240. 
The error estimation is described in the text. Two lines represent the best 
fit to the SW lobe and the NE lobe respectively. For the fitting procedure,
only points with SCP $ > 0$ of the each lobes were used. In the SW lobe, two 
interesting trends are visible. If we confirm the linear drop at the end of 
the CI track, with $\alpha_{\rm inj} \sim 0.82$, inspite of the large error of 
SCP and $\alpha_{\rm high}$, this can be interpreted as the CI bifurcation in 
Fig.~\ref{fig:sketch}. If this linear bifurcation is real, it implies that 
$B_{\rm sync}$ in the CI region is much stronger than $B_{\rm CMB}$.
Otherwise the CI bifurcation is not seen clearly.
The second points is the spread of data in the KP/JP range. Again inspite of 
the error bar, the trend visible here resembles the evolution of KP spectrum 
(see fig.~\ref{fig:B_CMB}), and $B_{\rm sync}$ may be several times stronger 
than $B_{\rm CMB}$.
}
\label{fig:DA240_scp_a}
\end{figure*}

Besides this asymmetry of the injection spectral indices, the two lobes reveal
quite different trends in the diagram. The synchrotron ageing theory, the CI 
model and the KP/JP model, well describes the trend in the SW lobe 
(Fig.~\ref{fig:DA240_scp_a}). The CI bifurcation is detected.
On the other hand, the majority of the SCP 
values in the NE lobe are well below zero. This is a clear case of spectral 
flattening. The remaining points with SCP $> 0$ are best fitted by the CI 
model. In the NE lobe $\alpha_{\rm high}$ commences with 0.5, then increases to
0.8 below SCP $ <0$. This is indicative of a non-relativistic strong shock as 
discussed in the last section. Since there is no clue of KP bifurcation (see 
Fig.~\ref{fig:scp}) or outreach of JP spectra on the SCP-$\alpha$ diagram, 
due to the sensitivity limit of the observation, we can not definitely prefer 
any model to the others, except for region CI mentioned in 
Fig.~\ref{fig:DA240_scp_a}.
Comparing Fig.~\ref{fig:DA240_scp_map} and Fig.~\ref{fig:DA240_scp_a},
it is found that the projected position of the CI bifurcation is the
channel of the brightness peak of the SW lobe to the SW extremity.
This possibly indicates that the physical condition of CI model, namely 
continuous injected electrons with no significant escape, is yet valid in 
this region.
\begin{figure*}[tbp]
  \centering\includegraphics[width=12cm]{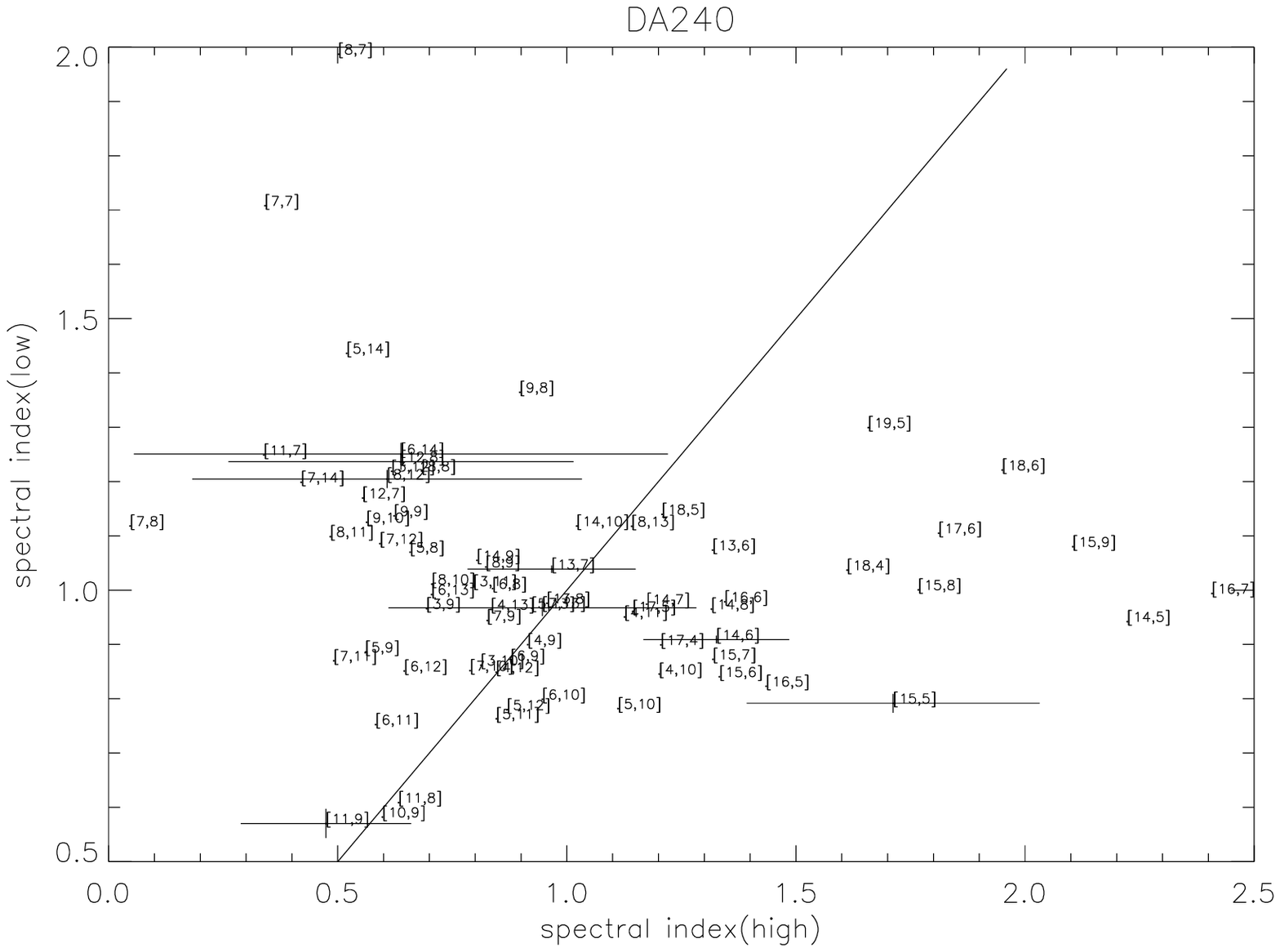}
  \caption[C-C diagram of DA\,240]{C-C diagram of DA\,240. The straight line 
corresponds to a pure power-law. The area to its left is populated by points
of spectral flattening, the area to its right contains points which show 
spectral steepening.
}
  \label{fig:DA240_a_a}
\end{figure*}
%
%
%
\subsubsection{NGC\,315}

\begin{figure*}[tbp]
  \centering\includegraphics[width=12cm]{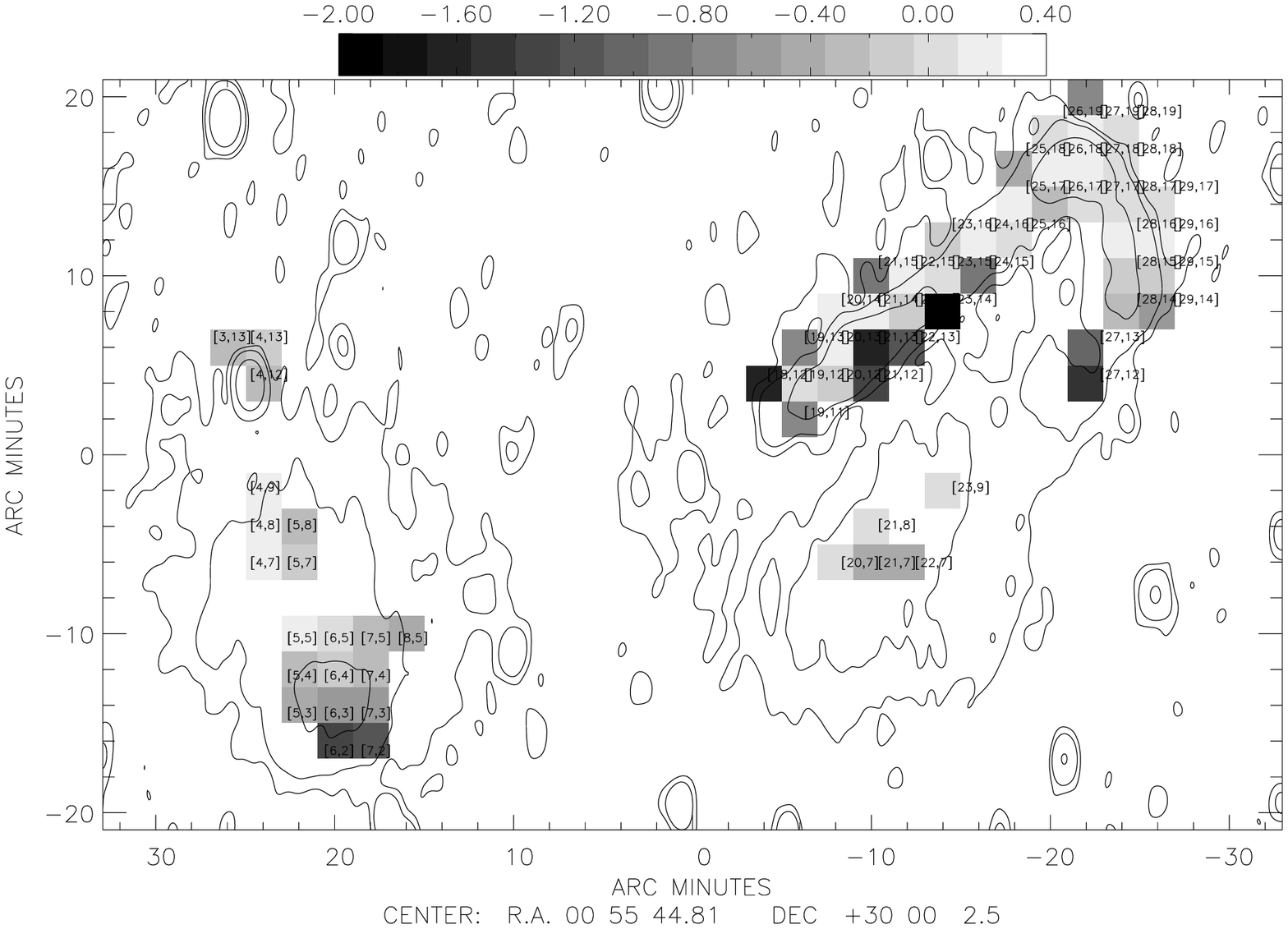}
  \caption[SCP map NGC\,31]{SCP map NGC\,315. 
Contours show the total intensities at 326~MHz \citet{mack:1997}. Contour 
levels are 3$\sigma_{\rm I}$, 10$\sigma_{\rm I}$ and 50$\sigma_{\rm I}$. 
Although its morphology is highly asymmetric, there is no significant spectral 
asymmetry along the major jet axis. Along the minor axis, in the SW to NE 
direction, the spectral curvature exhibits gradual steepening. This is clear 
in the whole SEern lobe and in the bow structure of NWern lobe. The relic tail 
of this structure also has flat spectral curvature.
}
  \label{fig:NGC315_scp_map}
\end{figure*}

Spectral flattening is present over the whole source. The value 
$\alpha_{\rm inj} \sim 0.58 $ in the NW lobe is consistent with the estimate 
of \citet{mack:1998}, who obtained 0.54 $\ldots$ 0.59. On the other hand, we 
cannot pro\-perly estimate the injection spectral index of the SE lobe, due to 
the small number of points with SCP $ >0$ and the large uncertainties. 
The general trend in the SE lobe 
implies a steep injection spectrum, $\alpha_{\rm inj} \sim 1.0$. The trends in 
the two lobes are neither symmetric nor asymmetric, but rather symmetric 
w.r.t the minor-axis (Fig.~\ref{fig:NGC315_scp_map}). At the southern ends 
of the two lobes, the spectral-upturning is striking. After that, towards 
the north, a gradual steepening follows.

The tracks of the NW and the SE lobe are well separated, which implies 
different injection indices, although the error is quite large. The reason 
for the extremely flat and even upturning curvature in the NW is unclear. 
Unresolved background sources or relativistic shocks could be the explanation.
\citet{ensslin:2001} suggest that the relic NW tail of NGC\,315 is 
re-accelerated by a cosmological shock wave. Our anal\-ysis demonstrates that 
the particles in both lobes have been re-accelerated. If the re-acceleration 
scenario is true, the spectral flattening implies that the energy threshold of 
this acceleration and/or $B_{\rm sync}$ of this region are higher than those 
of the injection spectrum. Spectral flattening plus an upturn are independent 
from possible missing short-spacing problems inherent to the 610~MHz WSRT data,
since it would be detectable via the Effelsberg single dish multi-frequency 
observations at 2.6, 4.8 and 10.6~GHz \citep{mack:1998} alone. However the 
prominence of the points with SCP $ < 0 $ could be correlated with the angular 
size of the 610~MHz data. It can be speculated that this is a viable 
explanation for the prominence of SCP $ < 0 $ in NGC315, which is by far the 
largest source in terms of angular size, $\Phi \sim 1\deg$. On the other hand, 
a value $\alpha$ of our SCP--$\alpha$ diagram is also obtained from 
single-dish 4.8 and 10.6~GHz data and shows $\alpha < \alpha_{\rm inj}$. 
The spectral flattening in NGC\,315 is mainly because of a low 
$\alpha_{\rm high}$.

\begin{figure*}[tbp]
  \centering\includegraphics[width=12cm]{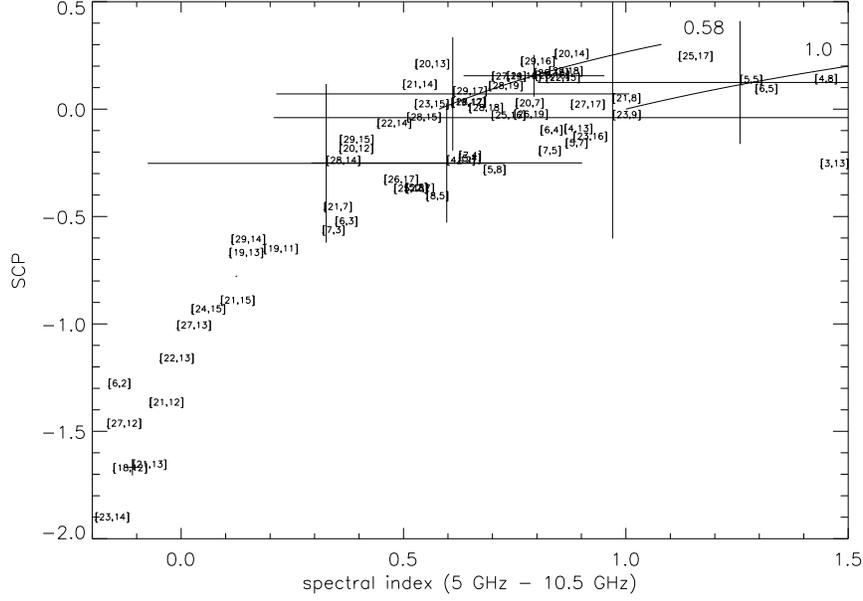}
  \caption[SCP-$\alpha$ diagram of NGC\,315]{SCP-$\alpha$ diagram of NGC\,315.
The SCP values are curiously flat in the source. The main jet to the NW end 
has a low injection index, $\alpha_{\rm inj} \sim 0.58$. $\alpha_{\rm inj}$ of 
the NWern relic tail and the SEern lobe is rather high, $\sim 1$. The NWern 
relic and the SEern lobe possibly have the same particle re-acceleration 
history.
}
  \label{fig:NGC315_scp_a}
\end{figure*}
\begin{figure*}[tbp]
  \centering\includegraphics[width=12cm]{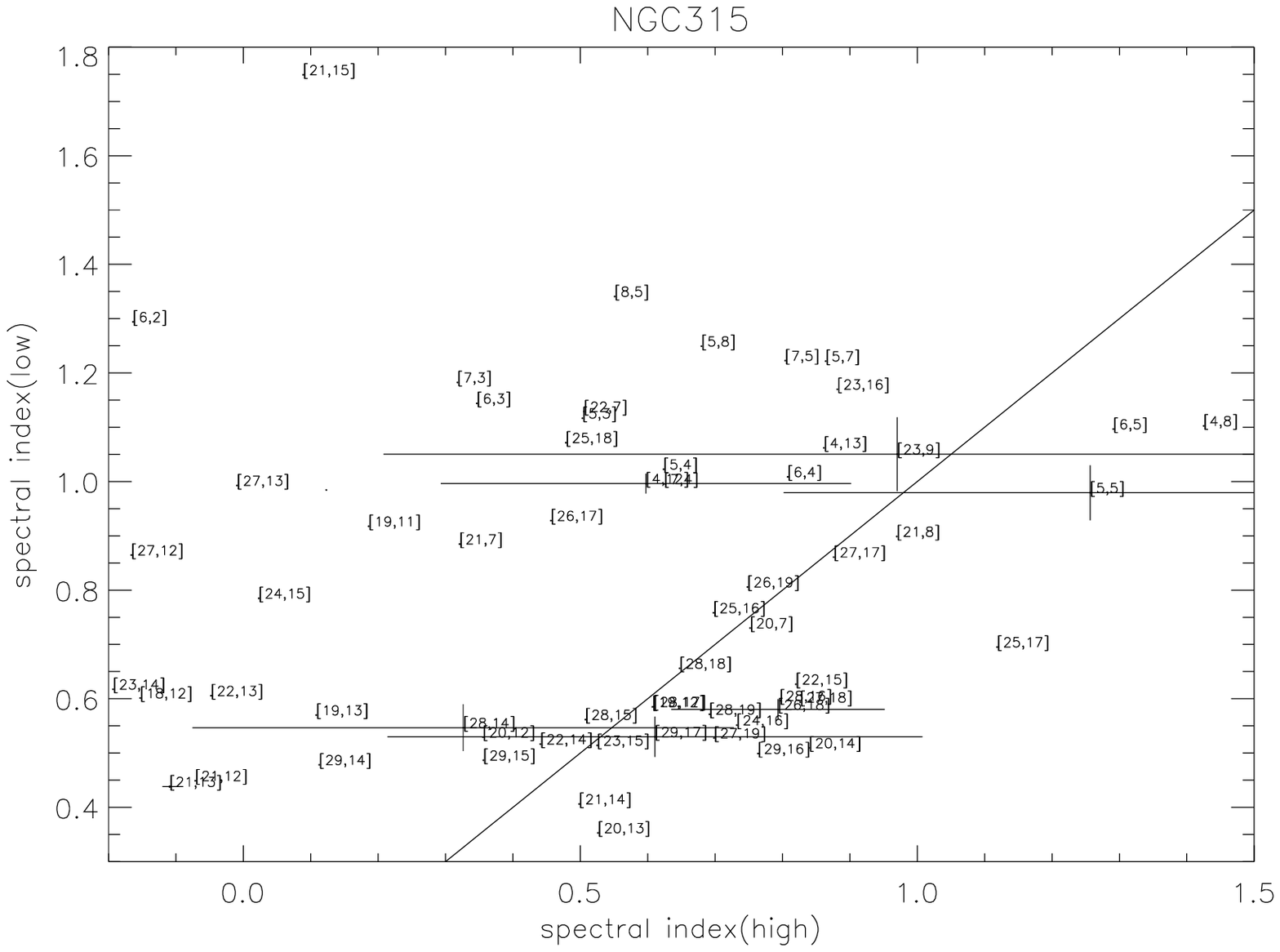}
  \caption[C-C diagram of NGC\,315]{C-C diagram of NGC\,315. See 
Fig.~\ref{fig:NGC315_a_a}.}
  \label{fig:NGC315_a_a}
\end{figure*}
%
%
\subsubsection{3C\,236}

This source is the largest known GRG, with $d \sim$~4.5 Mpc. It has a typical
FR II morphology. Our value of $\alpha_{\rm inj}$ of 0.7 in the SE lobe 
(Fig. \ref{fig:3C236_scp_a})
is in the range of the integrated synchrotron ageing estimate of 0.5 to 0.7
by \citet{mack:1998}. In the NW lobe, we obtain a steep $\alpha_{\rm inj}$ of
1.10. The integrated spectral index from synchrotron ageing theory for this
lobe is 0.7.

The spectral flattening shown at the extremity of the SE lobe is due to a 
known background source \citep{mack:1997}. In the NW lobe, we find two 
different values of $\alpha_{\rm inj}$, without any significant flattening. One
of them is well fitted by a steep spectrum, $\alpha_{\rm inj} \sim 1.10$. The 
trend of these data clearly 
shows the synchrotron ageing in the lobe. In the backflow or so-called bridge 
region, $\alpha_{\rm inj} \sim 0.7$ yields the best fit. Such an injection 
discrepancy between hot-spots and lobes has been reported for Cygnus A 
\citep{carilli:1991}. 
The injection spectral indices of the hot spots and lobes of Cygnus~A are 0.5 
and 0.75, respectively. The difference $\Delta \alpha_{\rm inj} \sim 0.4$ here 
is larger than in Cygnus~A. In Cygnus A, the hot-spots have 
$\Delta \alpha_{\rm inj} \sim 0.25$ flatter than their lobes, i.e. their 
backflows. 3C\,236 has a steeper spectrum ($\alpha_{\rm inj}$) in the advancing
region. An explanation for this $\alpha_{\rm inj}$ discrepancy 
\citep{carilli:1996} has not been found so far.

\begin{figure*}[tbp]
  \centering\includegraphics[width=12cm]{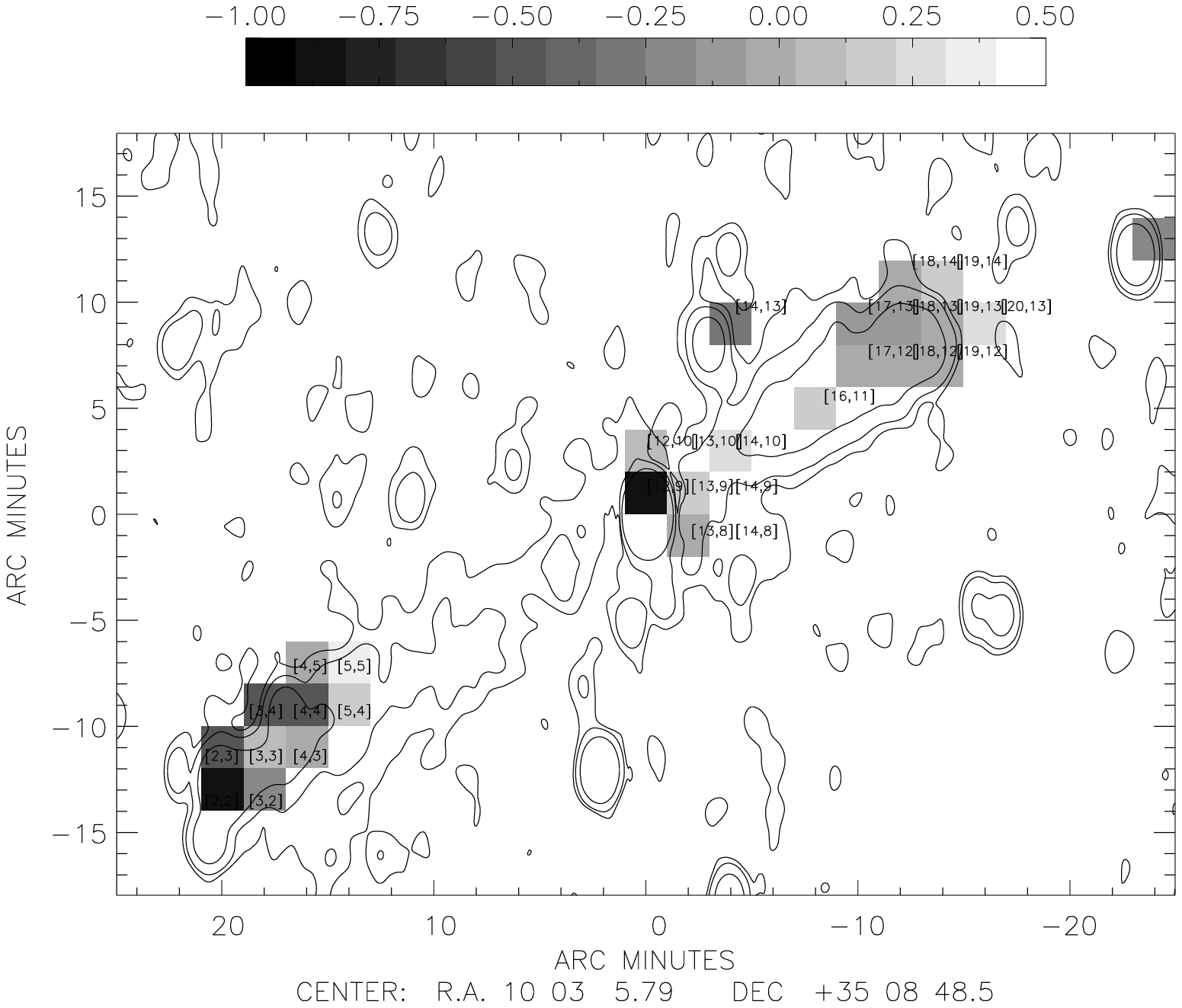}
   \caption[SCP diagram of 3C\,23]{SCP map of 3C\,236. 
The contours show the total intensities at 326~MHz \citet{mack:1997}.
Contour levels are 3$\sigma_{\rm I}$, 10$\sigma_{\rm I}$ and 
50$\sigma_{\rm I}$.
Two lines are the best fit results for the NW lobe and the SE lobe, 
respectively. The very flat SE extremity of SCP is due to a background source 
\citep{mack:1998}. The central core undergoes strong self-absorption, 
SCP $ < 0$. In general, SCP is symmetric in 3C\,236.
}
  \label{fig:3C236_scp_map}
\end{figure*}
\begin{figure*}[tbp]
  \centering\includegraphics[width=12cm]{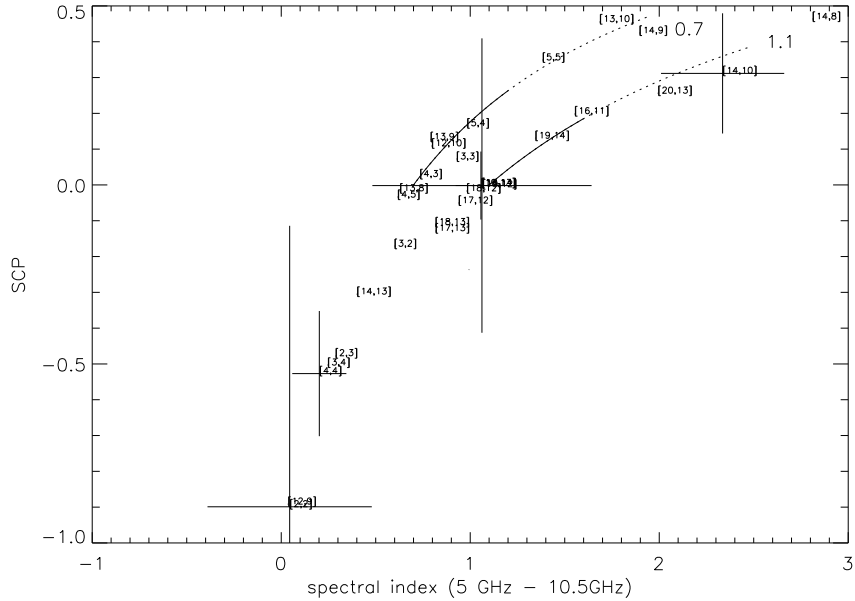}
  \caption[SCP-$\alpha$ diagram of 3C\,236]{SCP-$\alpha$ diagram of 3C\,236. 
Except for the central core and the background source, the points are well 
fit by two values of $\alpha_{\rm inj}$. The SE lobe has $\alpha_{\rm inj} 
\sim 0.7$. In the NW lobe, two values of $\alpha_{\rm inj}$s are seen. 
The near-to-core bridge has $\alpha_{\rm inj} \sim 0.7$, just like the SE lobe.
The NW outer lobe has a much steeper $\alpha_{\rm inj} \sim 1.1$.
}
  \label{fig:3C236_scp_a}
\end{figure*}
\begin{figure*}[tbp]
  \centering\includegraphics[width=12cm]{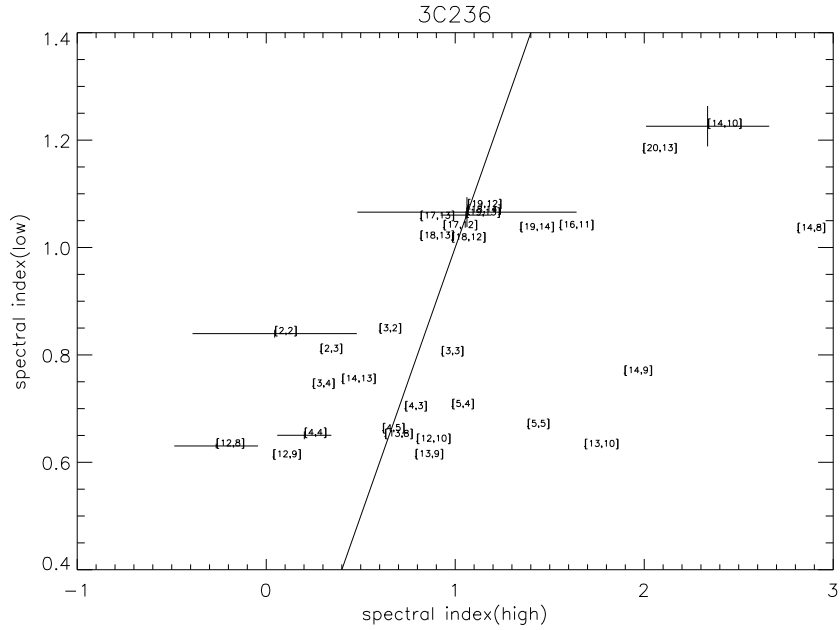}
  \caption[C-C diagram of 3C\,236]{C-C diagram of 3C\,236. 
See \ref{fig:DA240_a_a}.}
  \label{fig:3C236_a_a}
\end{figure*}

\subsection{CSS sources}
\begin{figure}[tbp]
  \resizebox{\hsize}{!}{\includegraphics{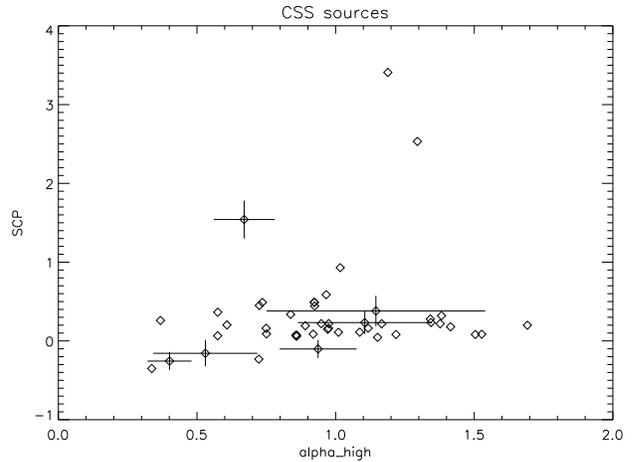}}
  \caption[47 CSS sources on the SCP-$\alpha$ diagram]{SCP-$\alpha$ diagram 
for 47 CSS sources. 40 sources have SCP values between 0 and 1. SCP $> 1$ 
clearly implies synchrotron self-absorption (See Fig.~\ref{fig:sketch}) at 
low frequencies. 3 out of 47 sources have SCP $> 1$. Due to the apparent 
synchrotron self-absorption at low frequencies, these sources were excluded
from the further analysis in this work.}
  \label{fig:css_scp}
\end{figure}
SCP-$\alpha$ diagrams can also be used for the analysis of samples of sources
for which only integrated flux densities are given. As an example we present 
the application of our method to a sample of 47 Compact Steep Spectrum (CSS) 
sources. \citet{murgia:1999} who have analyzed the flux densities of a sample 
of CSS sources in a profound synchrotron ageing study show that these sources 
have moderate spectral steepening, i.e. a difference of 
$\Delta \alpha \sim 0.5$
between low- and high-frequency spectral indices, which is predicted by the 
continuous injection model (CI). We have used this sample to test the SCP 
performance, which provides an alternative method for a quick analysis of 
synchrotron spectra. In essence, four frequencies 408 (327) MHz, 1.4GHz, 4.9 
(5.0) GHz and 10.7 (10.6, 8.1) GHz, were used. When no data were available at 
these frequencies, the total intensity at the frequencies given above in the 
brackets were taken. 

The points are found in the region where $\alpha_{\rm high}$ does not exceed
$\alpha_{\rm inj} + 0.5$. The obtained range of $\alpha_{\rm inj}$ values is 
rather wide. This confirms the results of \citet{murgia:1999}. The fact that
$\alpha_{\rm inj}$s shows more scatter in the SCP-$\alpha$ analysis than in 
the synchrotron model estimation is due the effect synchrotron self-absorption.
There is a clear trend that the sources with stronger $B_{\rm eq}$ and with 
smaller projected sizes have flatter $\alpha_{\rm inj}$. These are the compact 
GHz-Peaked Spectrum (GPS) source candidates, since due to their extreme 
compactness $d \leq 1 {\rm kpc}$, synchrotron self-absorption is liekly to be 
effective. The sample does not shows any correlation with redshift. This 
implies that the intrinsic magnetic fields proponder by far over the magnetic 
field equivalent to the cosmic microwave background.

\begin{figure}[tbp]
  \resizebox{\hsize}{!}{\includegraphics{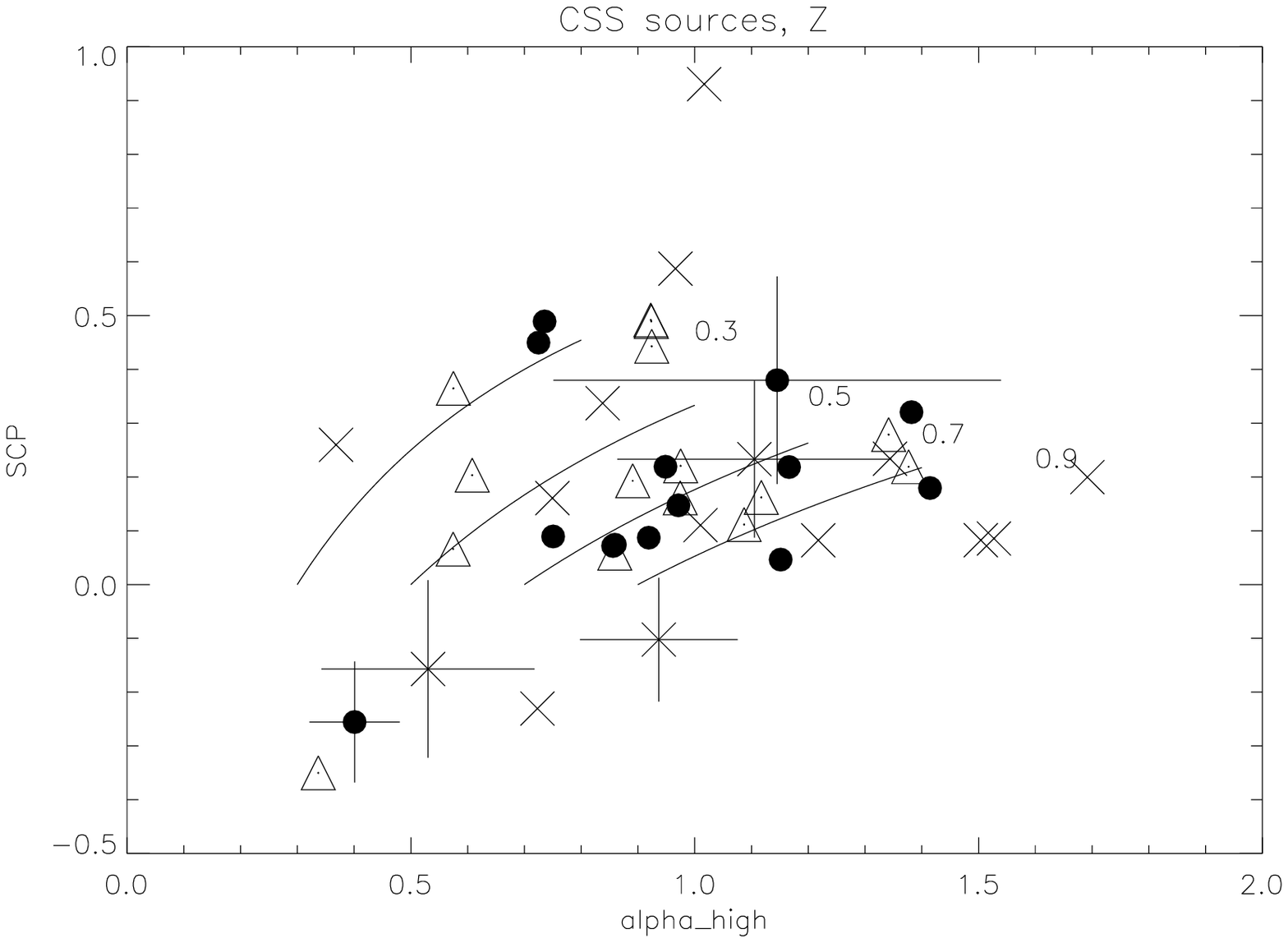}}
  \caption[CSS sources, redshift]{Filled Circles are CSS sources with 
$z < 0.5$. Open triangles are CSS sources with $0.5 \leq z < 1.0$. 
Crosses are CSS sources with $z \geq 1.0$. The sample does not show 
any z-related trend in the SCP-$\alpha$ diagram.}
  \label{fig:css_Z}
\end{figure}
\begin{figure}[tbp]
  \resizebox{\hsize}{!}{\includegraphics{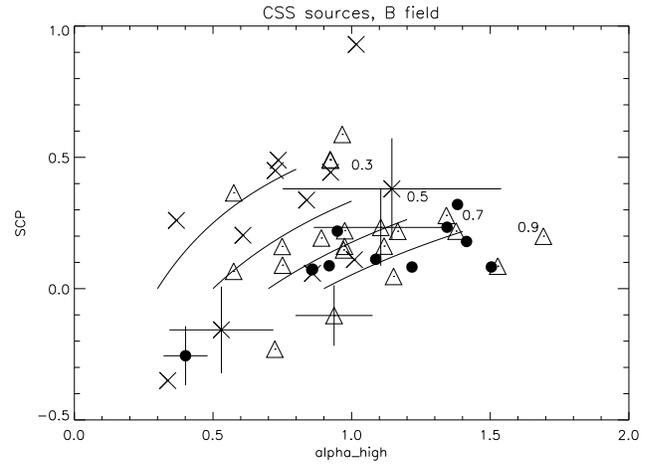}}
  \caption[CSS sources, $B_{\rm eq}$]{Filled Circles are CSS sources with 
$B_{\rm eq} < 5 \cdot 10^2 \mu G$. Open triangles are CSS sources with $5 
\cdot 10^2 {\rm \mu G} \leq B_{\rm eq} < 10^3 {\rm \mu G}$. Crosses are CSS 
sources with $B_{\rm eq} \geq 10^3 {\rm \mu G}$. $B_{\rm eq} \geq 10^3 {\rm \mu G}$ 
sources have flatter spectral curvature than the weaker $B_{\rm eq}$ CSS sources. 
$B_{\rm eq} \sim 10^3 {\rm \mu G}$ is typical value for GPS sources. GPS sources 
have their turn-over (due to synchrotron self-absorption) at GHz frequencies. Their 
$\alpha_{\rm low}$s are flat, $\alpha_{\rm low} < \alpha_{\rm inj}$, since they are 
estimated at $< 1$ GHz. The lines in the diagram were drawn assuming 
$\alpha_{\rm low} \geq \alpha_{\rm inj}$. As the result GPS sources tend to have 
flat $\alpha_{\rm inj}$, $< 0.5$. Some CSS sources have flat $\alpha_{\rm inj}$
\citep{murgia:1999} indeed, but these are not directly related to GPS 
sources, and none of them has extremely flat $\alpha_{\rm inj} < 0.3$.}
  \label{fig:css_B}
\end{figure}
\begin{figure}[tbp]
  \resizebox{\hsize}{!}{\includegraphics{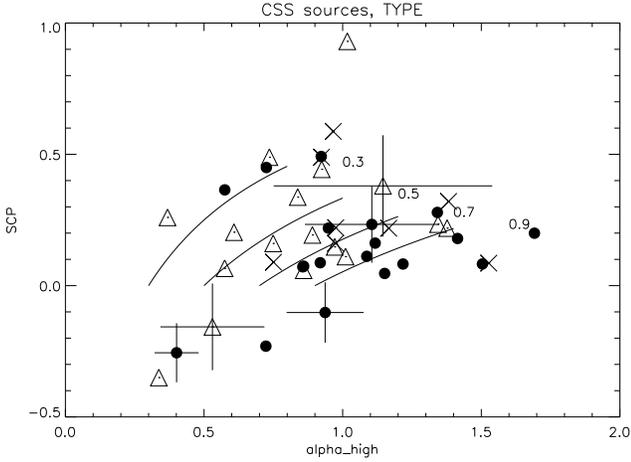}}
  \caption[CSS sources, source type]{Filled Circles are {\it lobe dominated} 
CSS sources. Open triangles are {\it core dominated} CSS sources. Crosses are 
{\bf uncertain} types of CSS sources. There is no clear trend to distinguish 
the three classes in the diagram. It can be partly due to the fact that CSS 
sources and GPS sources are not a proper definition of source morphology,
but of rather represent an evolutonary stage (visible in their spectrum). 
Relatively nearby GPS sources can be resolved and defined as {\it lobe dominated}, 
while distant CSS sources can be unresolved and defined as 
{\it core dominated}. Alternatively, some 'frustration scenario' could be 
working. A definite answer would be only possible with the improvement of
VLBI imaging.}
  \label{fig:css_TYPE}
\end{figure}
\begin{figure}[tbp]
  \resizebox{\hsize}{!}{\includegraphics{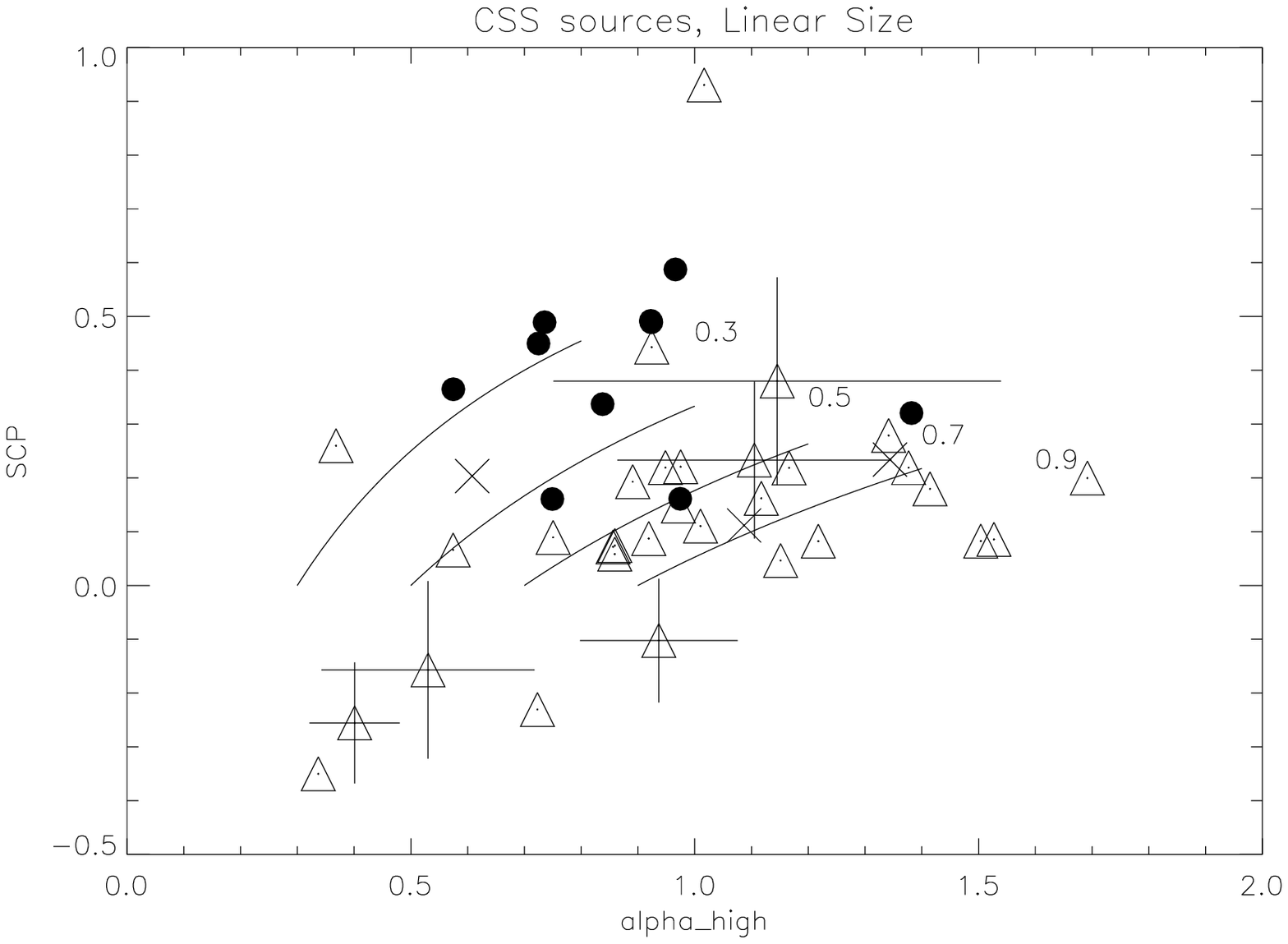}}
  \caption[CSS sources, length]{Filled Circles are $d < 1 \, {\rm kpc}$ sources. 
Open triangles are $1 \leq d < 10 \, {\rm kpc}$ sources. Crosses are $d \geq 10 
\, {\rm kpc}$ sources. Projected linear size, $d$, classes exhibit clear grouping 
of CSS sources as in $B_{eq}$ class. GPS sources are typically smaller than 
$1 \, {\rm kpc}$.}
  \label{fig:css_LS}
\end{figure}

\section{Discussion}

The above analysis implies that all studied sources have a complex history.
Now we discuss some possibilities. Before attempting any physical 
interpretation, we check again whether $SCP \sim -1$ could be generated by 
the possible missing short-spacing at 610~MHz. If anything, the possible 
missing short spacing at 610~MHz will steepen our $\alpha_{\rm low}$, which is 
already $>0$. Then,\\
\mspace $SCP = \frac{\alpha_{\rm high} - \alpha_{\rm low}}{\alpha_{\rm high} + 
\alpha_{\rm low}} \sim -1$, $\alpha_{\rm low} > 0$.\vspace{3mm}\\
This is possible, if $\alpha_{\rm low} \gg \alpha_{\rm high}$ or $\alpha_{\rm high} 
\sim 0$. $\alpha_{\rm low} \gg \alpha_{\rm high}$ is not known. The $SCP < 0$ trend 
can be emphasized, when $\alpha_{\rm low} > \alpha_{\rm high}$ and 
$\alpha_{\rm high} < 1$ are working together. But again the measured 
$\alpha_{\rm low}$ alone cannot explain the flattening up to $SCP \sim -1$. 
The trends in the sources are mainly due to the high-frequency curvature, 
$\alpha_{\rm high}$.

\subsection{Expansion loss, Energy Cut--off and Synchrotron Self--absorption}
Adiabatic expansion losses may play an important r\^{o}le as an energy
loss process of synchrotron sources. However, as \citet{carilli:1991} already
pointed out, although adiabatic expansion will shift the spectral break to
lower frequencies, the expansion does not change the spectral curvature.
Therefore, expansion losses will not affect the tracks of SCP--$\alpha$.

At low frequencies, there are also other physical processes that give rise to
spectral curvature, such as spectral turn--overs by synchrotron self--absorption 
in regions of high particle densities, or by a low--energy cut--off in the 
particle distribution. In the SCP--$\alpha_{\rm high}$ plane 
(Fig.~\ref{fig:sketch}), the low-frequency turn-over produces SCP $> 1$, which 
cannot be produced by any ageing processes. Strong self--absorption can even 
produce SCP $< 0$ and will be important in the central core regions, 
if $|\alpha_{\rm low}| > |\alpha_{\rm high}|$. Since $\alpha_{\rm low}$ will 
eventually approach $- 5/2$ in the Rayleigh--Jeans limit, this will be possible.

\subsection{Re-acceleration}
All three GRGs exhibit spectral flattening in some parts. In particular, 
NGC\,315 even shows signs of a spectral up--turn at high frequencies, and the 
majority of SCP points is under 0. Let us consider the case where the 
power-law injection spectrum is already established,\\
\mspace $N(E) \propto E^{-p}, \alpha_{\rm inj} = (p -1)/2$\vspace{3mm}\\
and where Fermi acceleration is working. By the Fermi process, the particles 
in each energy bin will be re-accelerated such as to yield a power-law of the 
form $N(E) \propto E^{-q}$. For the non-relativistic strong shock, $q = 2$. 
The final shape of these two power laws is described by the following 
integration:\\
%
\mspace $N(E) \propto E^{-q} \int^{E}_{E_0} {E'}^{-p} {E'}^{q-1} d{E'}, \, q>1, p>1$\vspace{3mm}\\
%
where ${E_0}$ is low energy cut off. This can be approximated 
\citep{blandford:1987, eilek:1991, sohn:2003}.
\begin{itemize}
\item[(i)] for $q < p$, $N(E) \propto E^{-q}$, $\alpha = (q - 1)/2$
\item[(ii)] for $p < q$, $N(E) \propto E^{-p}$, $\alpha = (p - 1)/2$
\end{itemize}
In case (i), spectral flattening and $SCP < 0$ is expected 
(Fig.~\ref{fig:sketch}).
An interesting result is that not every effective Fermi process results in a 
spectral flattening. In case (ii), the source will just look younger than
indicated by its kinematic age, inferred from the shift of the break frequency
towards higher frequencies \citep{parma:1999}. There is no flattening, since
the energy distribution of the re-accelerated particles follow $N(E) \propto
E^{-p}$, not $E^{-q}$.

Since GRGs are extraordinary extended, they should have a weak magnetic field,
 $\le B_{\rm CMB}$ \citep{mack:1998} and/or undergo re--acceleration processes 
during in their lifetime.

Considering the confusion of spectra of different components with different
spectral indices the observed high frequency spectral flattening indicates
that the flatter spectrum component is younger and secondary, i.e.
re--accelerated.
Otherwise we would not see the high frequency flattening, if the flatter 
spectrum component is as old as the steeper spectrum component. 
Or if the flatter spectrum component is dominant,  then we would
see only the flatter spectrum component in the radio frequency range and
then there would be no spectral flattening.

\subsection{(Equivalent) Magnetic fields}
The magnetic field $B_{\rm sync}$ and the equivalent field of the cosmic microwave
background, $B_{\rm CMB}$ determine the curvature beyond $\nu_{\rm br}$. In some 
models (e.g. \citet{eilek:1996}) magnetic fields produce a power-law spectrum 
when they are ordered in a power-law form. However, we restrict our discussion
to the curvature beyond $\nu_{\rm br}$, and to a simple homogeneous magnetic field
$B_{\rm sync}$ plus $B_{\rm CMB}$. Many radio galaxies as well as GRGs have weak 
magnetic fields \citep{feretti:1998, mack:1998, parma:1999}, assuming that the
equipartition estimation yields the strength of the magnetic field of the 
radiation region, $B_{\rm eq} = B_{\rm sync}$. The JP model becomes more appealing 
since it allows for pitch angle isotropization on a much shorter time scale than 
the radiation lifetime, $\tau_{\rm iso} \ll \tau_{\rm sync}$. However, KP `like' 
spectra are observed (e.g. \citet{carilli:1991}). In order to explain such KP 
spectra, variable $B_{\rm sync}$ fields were introduced 
\citep{tribble:1993, eilek:1997}.
In Tribble's model, the magnetic field has a Maxwellian distribution, while
in Eilek's model, magnetic fields are filamented, therefore have approximately
two components, $B_{\rm strong}$ and $B_{\rm weak}$. But again, any of these models 
requires that some portion of their $B$ fields is stronger than the equivalent
$B_{\rm CMB}$ to such as to produce the KP-like spectrum.

In any case, KP--like spectra are only possible when there is a strong 
magnetic field, with respect to $B_{\rm CMB}$. Therefore, the existence of KP 
spectra indicates that synchrotron radiation is the most important energy 
loss process in the region considered here. Furthermore, the variation of 
$B_{\rm sync}$, if any, will cause a broadening of the spectral turn-over at low 
frequencies.
%
%
%
%
%
%
\begin{figure}[tbp]
  \resizebox{\hsize}{!}{\includegraphics{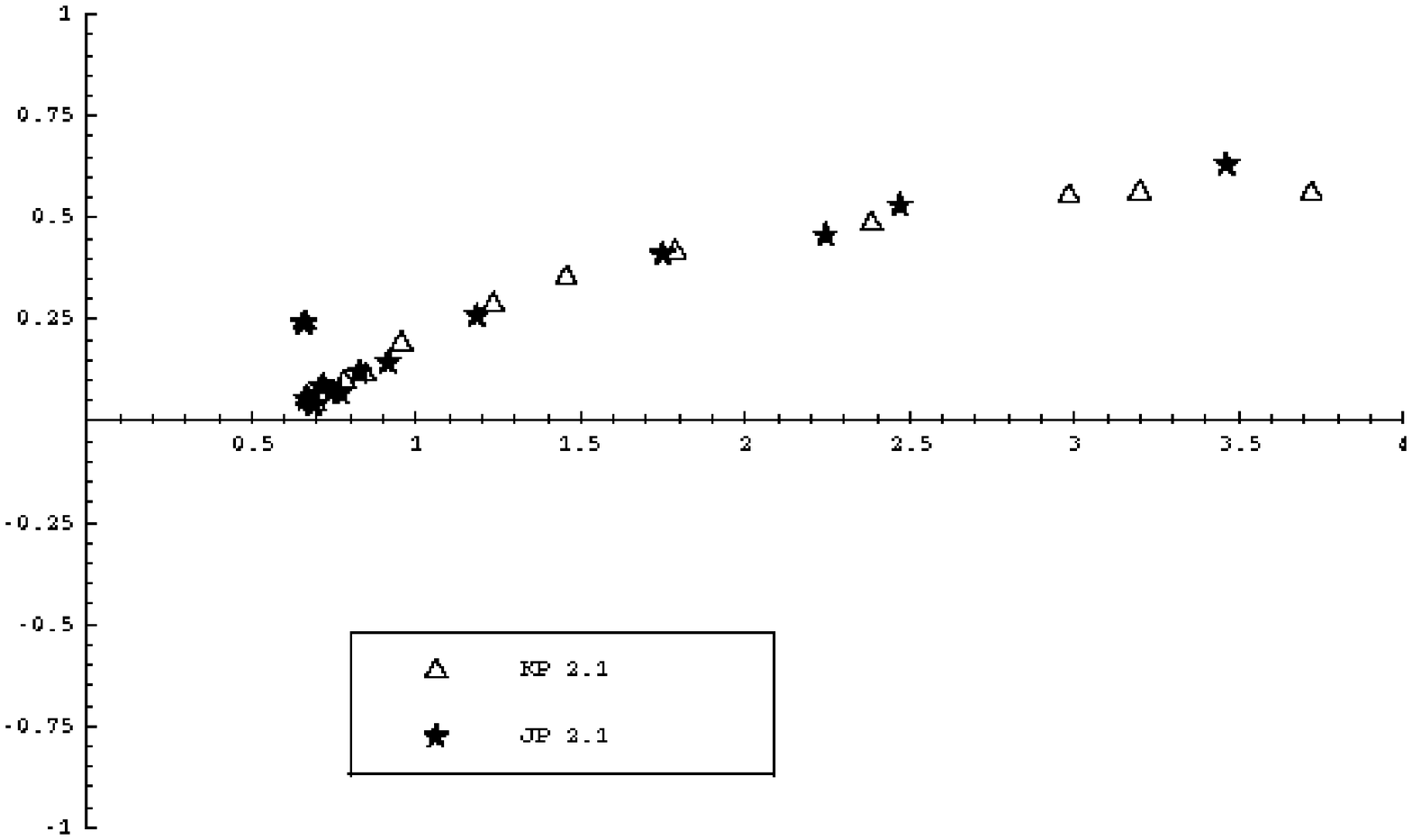}}
  \resizebox{\hsize}{!}{\includegraphics{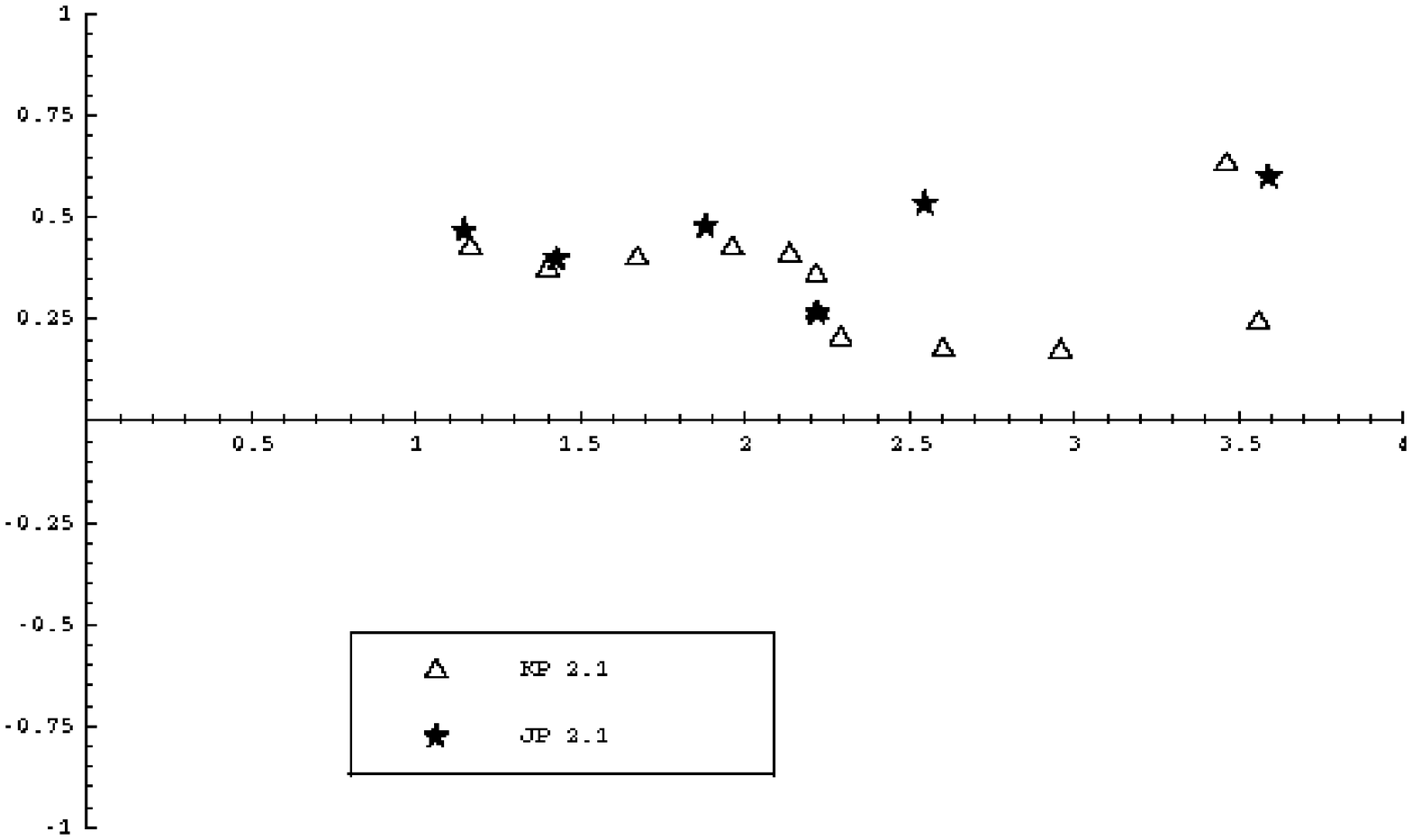}}
  \caption[Comparison of KP and JP spectra]{Comparison of KP and JP spectra 
under $B_{\rm CMB} = 3{\rm \mu G}$.The abscissa is $\alpha_{\rm high}$ and the 
ordinate is SCP. The frequency intervals were selected to be similar to the 
observating frequencies of the GRGs. In the simulation synchrotron losses, 
$IC_{\rm CMB}$ losses and slight energy cut--off effects are considered. From top 
to bottom, the magnetic field $B_{\rm sync}$ is given as $6 {\rm \mu G}$ and 
$15{\rm \mu G}$. Age, power index, $p = 2.1$($\alpha_{\rm inj} = 0.55$), and 
$B_{\rm CMB} = 3.2{\rm \mu G}$ are the same in all three diagrams. The pitch angle 
argument which makes the KP spectrum and the JP spectrum different is only valid 
when $B_{\rm sync}$ is distinctively stronger than $B_{\rm CMB}$. The bifurcation 
predicted in Fig.~\ref{fig:scp} is seen in the $B = 15{\rm \mu G}$ diagram, while 
beyond that KP also has an asymptotic tail, since in the end the effect of 
$B_{\rm CMB}$ appears. This will happen at a very steep $\alpha_{\rm high} > 2.5$.
}
  \label{fig:B_CMB}
\end{figure}
\section{Summary and conclusions}

We have investigated an alternative and very efficient method for the 
analysis of synchrotron spectra. We apply it both, to extended sources (like 
GRGs) and to the integrated flux densities of a sample of CSS sources. For 
all of them a thorough synchrotron ageing study has been performed which can 
be used for comparison. The information obtained from the spectral curvatures 
is manifold. 
The hot spots and jets possess pure power-law spectra, with particle ageing 
as expected. The spectral curvatures of the lobes exhibit both, spectral 
steepening and flattening. 

In DA\,240, there are CI spectra at the SW extremity, while KP/JP spectra show
up around the bright core of the SW lobe. We cannot find any bifurcation in 
the diagram, which serves as the definite distinction between the KP model 
and the JP model. More sensitive and/or higher-frequency observations are 
needed to reveal the bifurcation between the KP and JP models as shown 
between CI and KP/JP in DA\,240. If a KP bifurcation is seen, it can be 
interpreted as an identification of the existence of a strong $B_{\rm sync}$, 
i.e. $B_{\rm sync} > B_{\rm CMB}$.
As seen in Fig.~\ref{fig:B_CMB}, KP spectra can be identified. Otherwise, the 
spectra would look like JP spectra, due to the influence of the isotropic 
nature of $B_{\rm CMB}$.

The high-frequency spectral indices start at values of around 0.5, which is
indicative of non-relativistic strong shock acceleration. A possible origin of
the shock could be the interaction of radio galaxies with their surrounding
IGM/ICM \citep[e.g.][]{ensslin:2001}. Adiabatic expansion, the other 
significant energy loss process, does not affect the SCP-$\alpha$ diagram. 
The results demonstrate that the SCP provides crucial parameters for the continuum 
spectrum of synchrotron radiation, without the more complex modeling. 

Three characteristics that we have found in GRGs are not yet explained. The 
first is the origin of the asymmetry of the injection spectra of the 
radio lobes. Second, the physical explanation of the systematic flattening of 
$\alpha_{\rm high}$ compared to $\alpha_{\rm inj}$ is unknown. Third, there is a 
critical change of the re-acceleration efficiency showing up in the low- and 
high-frequency regime. We will investigate the environments of GRGs to find 
the possible reason.

In conclusion, it can be stated that the SCP-$\alpha$ diagram proves to be an
efficient method to derive important properties of synchrotron spectra which
otherwise can be determined only with the much more complex synchrotron 
ageing analysis. The SCP-$\alpha$ diagram and SCP map are especially useful 
to analyze a large number of sources and a large number of spectral points 
in a source. In those cases, the complex spectral analysis will give better 
estimation. However, this alternative tool provides fast estimates without 
losing accuracy significantly and provides an overview which is important
to understand synchrotron sources. Compared to the C-C diagram, 
the SCP-$\alpha$ diagram extracts injection spectral indices and possible 
synchrotron ageing models in a source more efficiently.
\begin{acknowledgements}
BWS is grateful to Heino Falke for discussions of the various shock 
acceleration conditions. KHM was supported by a Marie-Curie Fellowship. The 
authors are grateful to the anonymous referee for her/his fruitful comments.
\end{acknowledgements}
\bibliographystyle{apj}

\begin{thebibliography}{25}
\expandafter\ifx\csname natexlab\endcsname\relax\def\natexlab#1{#1}\fi

\bibitem[{Alexander(1987)}]{alexander2:1987}
Alexander, P. 1987, MNRAS, 225, 27

\bibitem[{Alexander \& Leahy(1987)}]{alexander:1987}
Alexander, P. \& Leahy, J. 1987, MNRAS, 225, 1

\bibitem[{Blandford \& Eichler(1987)}]{blandford:1987}
Blandford, R. \& Eichler, D. 1987, PhR, 154, 1

\bibitem[{Carilli \& Barthel(1996)}]{carilli:1996}
Carilli, C. \& Barthel, P. 1996, A\&A Rev., 7, 1

\bibitem[{Carilli {et~al.}(1991)Carilli, Perley, Dreher, \&
  Leahy}]{carilli:1991}
Carilli, C., Perley, R., Dreher, J., et al. 1991, ApJ, 383, 554

\bibitem[{Eilek \& Arendt(1996)}]{eilek:1996}
Eilek, J.~A. \& Arendt, P. 1996, ApJ, 457, 150

\bibitem[{Eilek \& Hughes(1991)}]{eilek:1991}
Eilek, J.~A. \& Hughes, P. 1991, {\it Beams and Jets in Astrophysics}
  (Cambridge Univ. Press)

\bibitem[{Eilek {et~al.}(1997)Eilek, Melrose, \& Walker}]{eilek:1997}
Eilek, J.~A., Melrose, D., \& Walker, M. 1997, ApJ, 483, 282

\bibitem[{En{\ss}lin {et~al.}(2001)En{\ss}lin, Simon, Biermann, Klein, Kohle,
  Kronberg, \& Mack}]{ensslin:2001}
En{\ss}lin, T., Simon, P., Biermann, P., et al. 2001, ApJ, 549, L39

\bibitem[{Feretti {et~al.}(1998)Feretti, Giovannini, Klein, Mack, Sijbring, \&
  Zech}]{feretti:1998}
Feretti, L., Giovannini, G., Klein, U., et al. 1998, A\&A, 331

\bibitem[{Jaffe \& Perola(1973)}]{jaffe:1973}
Jaffe, W. \& Perola, G. 1973, A\&A, 26, 423

\bibitem[{Kardashev(1962)}]{kardashev:1962}
Kardashev, N. 1962, Soviet Astr.-AJ., 6, 317

\bibitem[{Katz-Stone \& Rudnick(1994)}]{katz-ston:1994}
Katz-Stone, D. \& Rudnick, L. 1994, ApJ, 426, 116

\bibitem[{Katz-Stone \& Rudnick(1997)}]{katz-ston:1997}
---. 1997, ApJ, 488, 146

\bibitem[{Katz-Stone {et~al.}(1993)Katz-Stone, Rudnick, \&
  Anderson}]{katz-ston:1993}
Katz-Stone, D., Rudnick, L., \& Anderson, M. 1993, ApJ, 407, 549

\bibitem[{Klein {et~al.}(1995)Klein, Mack, Gregorini, \& Parma}]{klein:1995}
Klein, U., Mack, K.-H., Gregorini, L., et al. 1995, A\&A, 303, 427

\bibitem[{Mack {et~al.}(1997)Mack, Klein, O'Dea, \& A.G.}]{mack:1997}
Mack, K.-H., Klein, U., O'Dea, C., et al. 1997, A\&AS, 123, 423

\bibitem[{Mack {et~al.}(1998)Mack, Klein, O'Dea, Willis, \&
  Saripalli}]{mack:1998}
Mack, K.-H., Klein, U., O'Dea, C., et al. 1998, A\&A,
  329, 431

\bibitem[{Murgia {et~al.}(1999)Murgia, Fanti, Fanti, Gregorini, Klein, \&
  Mack}]{murgia:1999}
Murgia, M., Fanti, C., Fanti, R., et al. 1999, A\&A, 345, 769

\bibitem[{Pacholczyk(1970)}]{pacholczyk:1970}
Pacholczyk, A. 1970, {\it Radio Astrophysics} (Freeman, San Francisco)

\bibitem[{Pacholzcyk(1977)}]{pacholczyk:1977}
Pacholzcyk, A. 1977, {\it Radio Galaxies} (Pergamon press, Oxford)

\bibitem[{Park \& Schowengerdt(1983)}]{park:1983}
Park, S. \& Schowengerdt, R. 1983, Graphics \& Image Processing, 23, 256

\bibitem[{Parma {et~al.}(1999)Parma, Murgia, Morganti, Capetti, de~Ruiter, \&
  Fanti}]{parma:1999}
Parma, P., Murgia, M., Morganti, R., et al. 1999, A\&A, 344, 7

\bibitem[{Sohn(2003)}]{sohn:2003}
Sohn, B. 2003, PhD thesis, Univ. Bonn

\bibitem[{Tribble(1993)}]{tribble:1993}
Tribble, P. 1993, MNRAS, 261, 57

\end{thebibliography}

\end{document}